\documentclass[12pt]{article}

\usepackage{comment}
\usepackage[utf8]{inputenc}
\usepackage[margin=1in]{geometry}
\usepackage{amsmath}
\usepackage{multirow}
\usepackage{amssymb}
\usepackage{indentfirst}
\usepackage{amsfonts}
\newcommand\numberthis{\addtocounter{equation}{1}\tag{\theequation}}
\usepackage{scalerel}

\usepackage{natbib}
\usepackage{subcaption}
\usepackage[font=small, labelfont=bf]{caption}
\usepackage[colorlinks=true, allcolors=blue]{hyperref}
\usepackage[dvipsnames,svgnames]{xcolor}
\usepackage[english]{babel}
\usepackage[mathscr]{euscript}
\usepackage{rotating}

\usepackage{booktabs} 

\usepackage{fancyvrb} 

\usepackage{siunitx}

\usepackage{xpatch}
\xapptocmd\normalsize{%
 \abovedisplayskip=4pt plus 2pt minus 2pt
 \belowdisplayskip=4pt plus 2pt minus 2pt
}{}{}

\usepackage{thmtools}

\declaretheorem[
    name=Theorem,
    Refname={Theorem,Theorems}]{thm}

\declaretheorem[
    name=Assumption,
    Refname={Assumption,Assumptions}]{asu}
\declaretheorem[
    name=Remark,
    Refname={Remark,Remarks}]{rmk}

\newcommand{\loc}{\text{loc}}
\newcommand{\weekday}{\text{weekday}}
\newcommand{\sedentary}{\text{sedentary}}

\newcommand{\cee}{\text{CEE}}

\newcommand{\bA}{\bar{A}}
\newcommand{\ba}{\bar{a}}
\newcommand{\RR}{\mathbb{R}}
\newcommand{\PP}{\mathbb{P}}
\newcommand{\tp}{\tilde{p}}

\newcommand{\prob}{\text{Pr}}
\newcommand{\var}{\text{Var}}
\newcommand{\EE}{\mathbb{E}}

\usepackage{dsfont}

\def\spacingset#1{\renewcommand{\baselinestretch}%
{#1}\small\normalsize} \spacingset{1}

\begin{document}

  \title{\bf Modeling Time-Varying Effects of Mobile Health Interventions Using Longitudinal Functional Data from HeartSteps Micro-Randomized Trial}
   \date{\vspace{-5ex}} 
  \author{Jiaxin Yu \\ 
   Department of Statistics, University of California, Irvine\\ 
   Tianchen Qian \\
   Department of Statistics, University of California, Irvine}
\maketitle

\begin{abstract}
To optimize mobile health interventions and advance domain knowledge on intervention design, it is critical to understand how the intervention effect varies over time and with contextual information. This study aims to assess how a push notification suggesting physical activity influences individuals' step counts using data from the HeartSteps micro-randomized trial (MRT). The statistical challenges include the time-varying treatments and longitudinal functional step count measurements. We propose the first semiparametric causal excursion effect model with varying coefficients to model the time-varying effects within a decision point and across decision points in an MRT. The proposed model incorporates double time indices to accommodate the longitudinal functional outcome, enabling the assessment of time-varying effect moderation by contextual variables. We propose a two-stage causal effect estimator that is robust against a misspecified high-dimensional outcome regression nuisance model. We establish asymptotic theory and conduct simulation studies to validate the proposed estimator. Our analysis provides new insights into individuals' change in response profiles (such as how soon a response occurs) due to the activity suggestions, how such changes differ by the type of suggestions received, and how such changes depend on other contextual information such as being recently sedentary and the day being a weekday.
\end{abstract}

\noindent%
\vfill

\newpage

\spacingset{1.9} 

\section{Introduction}
\label{sec:introduction}

Mobile health (mHealth) interventions, such as push notifications delivered through smartphones and wearables, are being developed in various domains including physical activity, mental health, and substance abuse \citep{klasnja2018efficacy, battalio2021sense2stop, rabbi2018sara}. These interventions hold the promise to transform healthcare because they are affordable and accessible complements to in-person care. Micro-randomized trial (MRT) is the state-of-the-art experimental design for developing and optimizing such interventions \citep{klasnja2015microrandomized,dempsey2015randomised,qian2022microrandomized}. In an MRT, each individual is repeatedly randomized among treatment options hundreds or thousands of times, giving rise to longitudinal data with time-varying treatments. We will tackle the challenging task of inferring how mHealth interventions impact behavior and how such impact changes over time and depends on contexts, which are critical for developing and optimizing mHealth interventions and advancing domain knowledge.

This analysis is motivated by the HeartSteps study, an MRT for developing mHealth interventions for promoting physical activity among sedentary adults \citep{klasnja2015microrandomized}. We focus on assessing the time-varying effect of the activity suggestions, one of the intervention components in the HeartSteps app. For this component, each participant was randomized five times a day for 42 days, each time with probability 0.6 to receive a push notification suggesting either a brief walking activity or a stretching activity. The step count outcome was continuously measured via a wearable device. Previous research found a positive effect of the activity suggestion on an aggregate outcome, which is the total step count in a pre-specified 30-minute time window following each randomization \citep{klasnja2018efficacy}. However, important questions regarding how the intervention worked on a more granular level, such as how soon an individual responded to the activity suggestion and the impact of contextual factors like real-time location remain unanswered. We aim to answer such questions using the longitudinal functional step count data from HeartSteps MRT.

For analyzing MRT data, the primary quantity of interest is the causal excursion effect (CEE), which captures the difference in the outcomes between hypothetically following one intervention policy versus another for a specific time period. Existing methods for assessing CEE are all designed for longitudinal outcomes that are scalar at each time point \citep{boruvka2018assessing,qian2021estimating,shi2022assessing}. We build upon this literature and tackle the methodology challenge of modeling CEE on a longitudinal functional outcome.

Non-causal modeling of functional data has been studied extensively in the literature, with notable nonparametric function approximators such as spline-, kernel-, and reproducing kernel Hilbert space-based methods \citep{ruppert2003semiparametric, fan1996local, kimeldorf1971some}. The varying-coefficient model was proposed to model the nonlinear relationship between time-varying outcomes and predictors \citep{hastie1993varying, tan2012time, zhang2015varying}.

Nonparametric methods have also been used in causal modeling to fit nuisance parameters, such as propensity scores and outcome regressions, to improve robustness and efficiency of the causal parameter estimator \citep{hirano2003efficient, vanderlaan2011targeted, kennedy2016semiparametric, chernozhukov2018double}. In addition, causal effects themselves, such as the average causal effect conditional on a covariate vector, have been modeled nonparametrically \citep{robins2008higher,luedtke2016super,athey2016recursive,wager2018estimation,kennedy2020towards}. Researchers also used nonparametric methods to model causal effects with a continuous-valued treatment \citep{kennedy2017non, colangelo2020double}. To our knowledge, no nonparametric causal models have been proposed for longitudinal functional data.

We propose a time-varying effect model on CEE for MRT with longitudinal functional outcomes. The model incorporates two time indices, one for capturing how the causal effect varies over decision points in the MRT, and the other for capturing the causal effect on the functional outcome within each decision point. We use polynomial splines with B-spline basis to model the time-varying effects because of their computational efficiency and stability \citep{huang2004polynomial, hastie1987generalized}. We established the asymptotic property of the proposed estimator, including consistency, rate of convergence, and robustness to a misspecified high-dimensional nuisance working model. Simulations were conducted to validate the asymptotic property and an asymptotic variance formula.

We apply the method to assess the impact of the activity suggestions in the HeartSteps MRT. Our analysis reveals distinct time-varying effects of activity suggestions, which were dependent on key moderating factors like recent activity level and day type---individuals were generally more responsive on weekdays or when they were recently sedentary. We also find that responses to the activity suggestions began approximately 10 minutes after receiving a push notification and tapered off around the 40-minute mark. In addition, the two types of activity suggestions were found to have different declining patterns in their effectiveness over the study. These findings provided insights into how the activity suggestions affected an individual's behavior, the proper definition of the proximal outcome for future physical activity MRTs, and directions to improve the activity suggestions.

We note that longitudinal functional data is common in MRTs with wearable- or sensor-collected outcomes. Other examples include continuously measured stress levels in a smoking cessation MRT \citep{battalio2021sense2stop}, continuously measured app engagement in a drinking reduction MRT \citep{bell2023notifications}, and continuously measured calories burned in a physical activity MRT \citep{mishra2023text}. Therefore, our method can be applied to many problems.

The rest of the paper is organized as follows. We introduce the data and the research question of interest in Section \ref{sec: HeartSteps MRT Data}. We define the causal excursion effect in Section \ref{sec:notation}. The proposed model and the estimation procedure are described in Section \ref{sec:method}. Asymptotic theory is presented in Section \ref{sec:theory}. Section \ref{sec:simulation} presents simulation results, illustrating the finite sample performance of our estimation and inference procedures. Section \ref{sec:application} presents the analysis result using the HeartSteps MRT data. Section \ref{sec:discussion} concludes with discussions.

\section{HeartSteps MRT Data}
\label{sec: HeartSteps MRT Data}

\begin{figure}
    \centering
    \includegraphics[scale = 0.55]{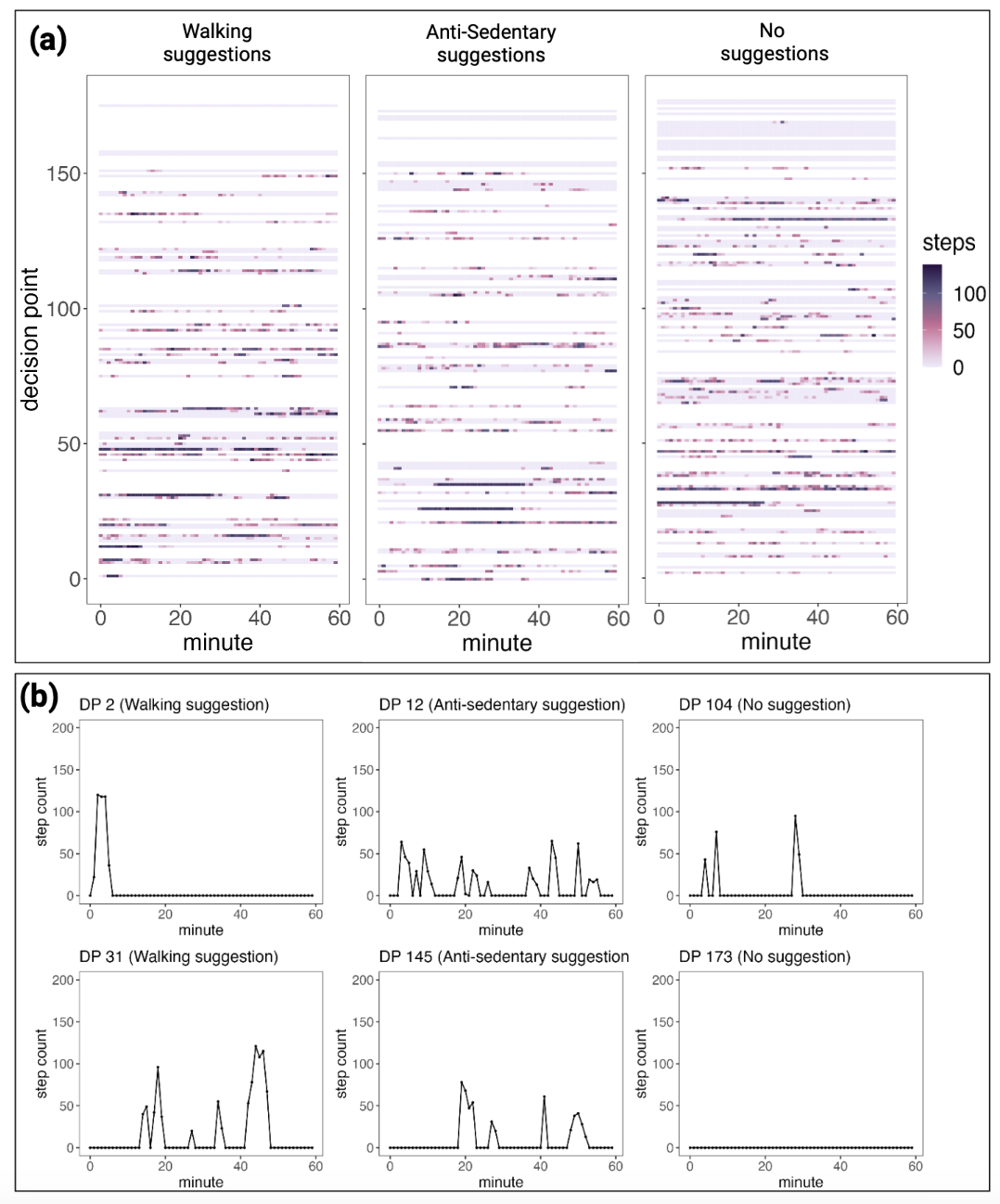}
    \caption{(a) The minute-level step count over the 60-minute window following each of the 210 decision points for one participant. Decision points with different types of suggestions are separated into three panels. Darker color indicates higher step count in that minute. (b) The minute-level step count from six selected decision points for that participant.}
    \label{fig:exploratory plot_heatmap}
\end{figure}

We use data from HeartSteps I (hereafter referred to as HeartSteps and can be accessed on \url{https://github.com/klasnja/HeartStepsV1/tree/main/data_files}), the first of a sequence HeartSteps MRTs, whose goal was to develop mHealth interventions to help sedentary adults achieve and maintain recommended levels of physical activity \citep{klasnja2015microrandomized}. The HeartSteps app featured multiple components, focusing on delivering contextually tailored activity suggestions through smartphone notifications to participants. These suggestions included \textit{walking suggestion} requiring 2-5 minutes to finish, \textit{anti-sedentary suggestion} prompting brief movements and requiring 1-2 minutes to finish, or no suggestion. Each of the 37 participants was randomized five times a day for 42 days, each time with probability 0.3, 0.3, and 0.4 to receive a walking suggestion, an anti-sedentary suggestion, or no suggestion, respectively. The randomization times were selected by participants at the beginning of the study and occurred at least 90 minutes apart. Each participant wore a wristband tracker which continuously recorded their minute-level step count. Figure \ref{fig:exploratory plot_heatmap} shows the step count data collected from a participant. As depicted in Figure \ref{fig:exploratory plot_heatmap}(a), the large variability in the data demonstrates that careful modeling is needed.

Previous research found that averaging over the 42 days, the activity suggestions had a positive effect on the total step count in a 30-minute window following randomization, and the walking suggestions resulted in a larger effect on this aggregate outcome than the anti-sedentary suggestions \citep{klasnja2018efficacy}. We will tackle several unanswered questions that are of both scientific interest and relevance for optimizing the activity suggestions: How soon does a participant respond to an activity suggestion? Does the effect of the activity suggestion decay linearly or nonlinearly over the study? Does the time-varying effect of the activity suggestion depend on contextual information, such as being recently sedentary and being on a weekday? And does the response vary between walking and anti-sedentary suggestions?

\section{Preliminaries}
\label{sec:notation}

\subsection{Notation and Observed Data}

The data includes observations from $n$ individuals, each observed at $J$ decision points where interventions are randomly assigned.  Let $X_{ij}$ be the individual's contextual information collected after decision point $j-1$ and up to decision point $j$. Let $A_{ij}$ be the treatment assignment at  decision point $j$. After treatment is assigned at decision point $j$, the proximal outcome is measured for a time window of length $T$. We denote the outcome by $Y_{ij} = (Y_{ij1},...,Y_{ijT})^T$, with each $Y_{ijt}$ being a scalar. We assume that $Y_{ij}$ is measured before decision point $j+1$. For clarity, the method and theory will be presented on the setting where $A_{ij}$ is binary with value 1 if subject $i$ was assigned treatment at decision point $j$ and 0 otherwise. The data analysis in Section \ref{sec:application} will employ a generalization of the method to categorical treatments, with derivations in Section D of the Supplementary Material.

In intervention design, researchers might consider it unsafe or unethical to send push notifications at certain decision points, such as during driving. This concept is termed ``(un)availablility'' in the literature \citep{boruvka2018assessing}. Formally, $X_{ij}$ includes an indicator $I_{ij}$, with $I_{ij} = 1$ denoting being available at decision point $j$ and $I_{ij} = 0$ otherwise. If $I_{ij} = 0$, then $A_{ij} = 0$ deterministically.

The overbar is used to denote a sequence of variables up to a decision point; for example, $\bA_{ij} = (A_{i1},..., A_{ij})$. Information accrued up to decision point $j$ is represented by the history $H_{ij} = ({X}_{i1}, {A}_{i1}, {Y}_{i 1},...,{X}_{i,j-1}, {A}_{i,j-1}, {Y}_{i, j-1}, {X}_{ij})= (\bar{X}_{ij}, \bA_{i, j-1}, \bar{Y}_{i, j-1})$. The randomization probability for $A_{ij}$ is a constant in HeartSteps but can generally depend on $H_{ij}$ and we denote it by $p_{ij}(H_{ij}) = P(A_{ij} = 1|H_{ij})$. We assume that the data from different individuals are independent and identically distributed samples from an unknown distribution. Throughout, we represent random variables or vectors with uppercase letters, and their realized values with lowercase letters.

In HeartSteps, each participant was in the study for $J = 5 \times 42 = 210$ decision points. $X_{ij}$ records information such as the individual's location and the current day of the week. $A_{ij}$ represents whether an activity suggestion is sent to individual $i$ at decision point $j$, randomized with a constant probability 0.6 when $I_{ij} = 1$. $I_{ij} = 0$ if the individual is driving or already walking. We examine the effect of the activity suggestion over the subsequent 60 minutes, thus $T = 60$ and $Y_{ij} = (Y_{ij1},...,Y_{ij60})^T$. 

We use $[n]$ to denote the set $\{1,2,\ldots,n\}$. For reference, a list of symbols used is summarized in Section F of the Supplementary Material.

\subsection{Potential Outcomes and Causal Effects}

To define causal effects, we use the potential outcomes framework \citep{rubin1974estimating, robins1986new}. Specifically, let $X_{ij}(\ba_{i,j-1})$ be the observation that would have been observed had that individual received the treatment sequence $\ba_{i, j-1}$. Let $Y_{ij}(\ba_{ij})$ be the collection of potential outcomes $\{Y_{ij1}(\ba_{ij}),...,Y_{ijT}(\ba_{ij})\}^T$ under treatment sequence $\ba_{ij}$.


mHealth interventions are typically designed to influence an individual proximally in time. For example, the activity suggestions in HeartSteps aimed to urge individuals to stand up and walk in the moment. We define the causal excursion effect (CEE) of the treatment $A_{ij}$ on the proximal outcome $Y_{ijt}$ to capture such impacts:
\begin{align}
    \cee \{j,t, S_{ij}(\bA_{i,j-1})\} = \EE\{Y_{ijt}(\bA_{i,j-1}, 1) - Y_{ijt}(\bA_{i,j-1}, 0) \mid S_{ij}(\bA_{i,j-1}), I_{ij}(\bA_{i,j-1}) = 1 \}, \numberthis
    \label{eq: causal effect}
\end{align}
where $S_{ij}(\bA_{i,j-1})$ is a vector of summary variables formed from $H_{ij}(\bA_{i,j-1})$ based on researchers' analysis goal.  Expression \eqref{eq: causal effect} represents the contrast between the expected outcomes at $(j, t)$ under two treatment policies that are viewed as ``excursions'': following the MRT policy until decision point $j-1$, then either receiving treatment at decision point $j$ or not receiving treatment at decision point $j$. The causal effect is marginal over the distribution (under the MRT policy) of $\bA_{i,j-1}$ not included in $S_{ij}(\bA_{i,j-1})$. The effect is conditional on the moderators of interest $S_{ij}(\bA_{i,j-1})$, and different choices of $S_{ij}(\bA_{i,j-1})$ can address different scientific questions. For example, setting $S_{ij}(\bA_{i,j-1}) = \emptyset$ yields a fully marginal effect. Setting $S_{ij}(\bA_{i,j-1}) = \loc_{ij}$ yields causal effect moderated by the real-time location of the participant, $\loc_{ij}$. The effect is conditional on availability $I_{ij}(\bA_{i,j-1}) = 1$; i.e., we only focus on available times, because it is unethical or impractical to send interventions when someone is unavailable. Expression \eqref{eq: causal effect} generalizes the CEE in \citet{boruvka2018assessing} and \citet{qian2021estimating} to longitudinal functional outcome: for a fixed decision point and a moderator value, $\cee(j,\cdot,S_{ij})$ is a function over $t$, i.e., a time-varying causal effect ``curve''.

\subsection{Identification}
\label{Identification}

To express the causal effect \eqref{eq: causal effect} in terms of the observed data, we assume the following:

\begin{asu}[Consistency]
    \label{assumption:consistency}
    The observed data is the same as the potential outcome under the observed treatment assignment. Specifically, $Y_{ij} = Y_{ij}(A_{ij})$ and $X_{ij} = X_{ij}(A_{ij-1})$ for $j \in [J]$.
\end{asu}

\begin{asu}[Positivity]
    \label{assumption:positivity}
     If $\prob(H_{ij} = h, I_{ij} = 1) > 0$, then $\prob(A_{ij} = a|H_{ij} = h, I_{ij} = 1) > 0$ for $a \in \{ 0, 1\}$ and $j \in [J]$.
\end{asu}
    
\begin{asu}[Sequential ignorability]
    \label{assumption:seqignorability}
    The potential outcomes $\{Y_{ij}(\ba_{ij})$, $X_{i,j+1}(\ba_{ij})$, $\ldots, Y_{iJ}(\ba_{iJ})\}$ are independent of $A_{ij}$ conditional on $H_{ij}$ for $j \in [J]$.
\end{asu}

In a micro-randomized trial, positivity and sequential ignorability are satisfied by design because the treatment is sequentially randomized with known, nonzero probabilities when the participant is available. The consistency assumption is valid when an intervention sent to one participant doesn't influence the outcomes of others. This is a reasonable assumption in HeartSteps, where participants do not interact with each other. We prove in Section E of the Supplementary Material that under these assumptions, the causal effect \eqref{eq: causal effect} can be written in terms of the observed data:
\begin{align*}
    \cee(j,t,S_{ij}) = \EE\{\EE(Y_{ijt} \mid A_{ij}=1, H_{ij}) - \EE(Y_{ijt} \mid A_{ij}=0,H_{ij}) \mid S_{ij}, I_{ij} = 1\}
\end{align*}


\section{A Time-Varying Causal Excursion Effect Model}
\label{sec:method}

\begin{asu}[Additive model]
    \label{assumption:Additive model}
    We consider an additive model on the causal excursion effect: 
    \begin{align*}
    \cee(j,t,S_{ij}) = f_0(S_{ij})^T \beta_0 + f_1(S_{ij})^T \beta_1(j) + f_2(S_{ij})^T \beta_2(t), \numberthis \label{eq:cee-model}
        \end{align*}
        where $f_0(S_{ij})$, $f_1(S_{ij})$, and $f_2(S_{ij})$ are three vectors of features that typically include an intercept, and $\beta_0$, $\beta_1(j)$, and $\beta_2(t)$ are the corresponding coefficients. Formally, for $r \in \{0, 1, 2\}$, $f_r(s) \in \RR^{L_r}$ is a vector of prespecified transformation of $s$, $\beta_0 \in \RR^{L_0}$ is an unknown parameter, $\beta_1(j)$ is an unknown function of index $j$ and takes value in $\RR^{L_1}$, and $\beta_2(t)$ is an unknown function of index $t$ and takes value in $\RR^{L_2}$. We assume $\beta_1(1) = 0$ and $\beta_2(1) = 0$ for identifiability.
\end{asu}

For example, when the researcher is interested in marginal effect and sets $S_{ij} = \emptyset$, model \eqref{eq:cee-model} becomes $\cee(j, t) = \beta_0 + \beta_1(j) + \beta_2(t)$, where $\beta_1(j)$ captures how the causal effect varies across decision points and $\beta_2(t)$ captures variation within a decision point. When the researcher is interested in effect moderation by real-time location and thus sets $S_{ij} = \loc_{ij}$, model \eqref{eq:cee-model} becomes
\begin{align*}
    \cee(j,t,\loc_{ij}) = \beta_{01} + \loc_{ij} \beta_{02} + \beta_{11}(j) + \loc_{ij} \beta_{12}(j) + \beta_{21}(t) + \loc_{ij} \beta_{22}(t).
\end{align*}
Terms like $\loc_{ij}\beta_{22}(t)$ are known as ``varying coefficient'' terms \citep{hastie1993varying}.

\begin{rmk}
This additive model may fail if $S_{ij}$'s transformation is incorrect or if the model isn't additive in time indices $j$ and $t$. For the first scenario, one may use data-adaptive methods to model $f_0(s), f_1(s)$, and $f_2(s)$  \citep{zhang2015varying}, although this was not a concern in our analysis for HeartSteps because the moderators we considered were binary. For the second scenario, nonparametric models that are more general than additive models such as tensor product basis may be considered \citep{he1996bivariate}.
\end{rmk}

We approximate $\beta_1(\cdot)$ and $\beta_2(\cdot)$ using basis expansion:
\begin{align}
    \beta_{1l}(j) \approx \sum_{k = 1}^{K_{1l}}\gamma^{(1)}_{lk}\phi_{lk}(j), \ \ \ \ \beta_{2l}(t) \approx \sum_{k = 1}^{K_{2l}}\gamma^{(2)}_{lk}\psi_{lk}(t),
    \label{eq:basis-expansion}
\end{align}
where $\{\phi_{lk}(j): k \in [K_{1l}]\}$ and $\{\psi_{lk}(t): k \in [K_{2l}]\}$ denote the basis functions on ${1,2,\ldots,J}$ and ${1,2,\ldots,T}$, respectively, and $\gamma^{(1)}_{lk}$ and $\gamma^{(2)}_{lk}$ are the corresponding spline coefficients. We use B-splines to approximate $\beta_1(\cdot)$ and $\beta_2(\cdot)$; For an overview of B-splines, see, for example, \citet{de1978practical} and \citet{huang2004polynomial}. 

We propose the following two-stage procedure for estimating $\beta_0$ and $\gamma^{(r)}_{lk}$. In Stage 1, a nuisance function $\eta(j, t, H_{ij}) = E(Y_{ijt} \mid H_{ij}, I_{ij} = 1)$ is fitted using nonparametric methods. In Stage 2, we estimate $\beta_0$ and $\gamma^{(r)}_{lk}$ by minimizing a weighted and centered least squares criteria function with $\hat\eta(j, t, H_{ij})$ from Stage 1 plugged in:
\begin{align}
    \sum_{i=1}^n \sum_{j=1}^J I_{ij} W_{ij} \sum_{t=1}^T \bigg[ Y_{ijt} & - \hat\eta(j, t, H_{ij}) - \{A_{ij} - \tp_{ij}(S_{ij}) \} \nonumber \\
    \times \Big\{f_0(S_{ij})^T \beta_0 + & \sum_{l}^{L_1}f_{1l}(S_{ij}) \sum_{k}^{K_{1l}}\gamma_{lk}^{(1)}\phi_{lk}(j) + \sum_l^{L_2}f_{2l}(S_{ij}) \sum_{k}^{K_{2l}}\gamma_{lk}^{(2)}\psi_{lk}(t) \Big\}  \bigg]^2. \label{eq:estimating-equation}
\end{align}
Here, $W_{ij} = A_{ij} \tp_{ij}(S_{ij}) / p_{ij}(H_{ij})+ (1-A_{ij})\{1-\tp_{ij}(S_{ij})\} / \{1-p_{ij}(H_{ij})\}$
is a marginalization weight also used in other semiparametric estimators for causal excursion effects \citep{boruvka2018assessing,qian2021estimating}, and the numerator probability $\tp_{ij}(S_{ij}) \in (0,1)$ can be chosen arbitrarily as long as it only depends on $S_{ij}$. In HeartSteps, one can simply set $\tp_{ij}(S_{ij}) = 0.6$. The resulting estimators for $\beta_1(\cdot)$ and $\beta_2(\cdot)$ are
\begin{align*}
    \hat\beta_{1l}(j) = \sum_{k = 1}^{K_{1l}}\hat\gamma^{(1)}_{lk}\phi_{lk}(j); \ \ \ \ \hat\beta_{2l}(t) = \sum_{k = 1}^{K_{2l}}\hat\gamma^{(2)}_{lk}\psi_{lk}(t).
\end{align*}

As will be established in Section \ref{sec:theory}, consistency of $\hat\beta$ is agnostic to the nonparametrically estimated nuisance function $\hat\eta(j, t, H_{ij})$ in Stage 1, in terms of both the covariates included in the estimator and the choice of the nonparametric method. This desirable robustness property, which is a consequence of the centering and the weighting, protects against misspecification of the potentially complex $\eta(j, t, H_{ij})$ due to the high-dimensional $H_{ij}$.

\subsection{Selection of Smoothing Parameter}
\label{subsec:CV}

 We use B-splines with equally spaced knots, and we select the number of knots $K_{rl}$ data-adaptively using $K$-fold cross-validation. Specifically, we split the subjects into $K$ groups with roughly equal sizes. Let $k[i]$ be the group that subject $i$ is in. Let $\hat{\beta}_{r}^{(-k[i])}(\cdot), r = 1, 2$ be the spline estimator and $\hat\beta_0^{(-k[i])}$ be the intercept estimator obtained from the data with the observations from $k[i]$ group removed. Then, similar to \citet{huang2004polynomial}, we define the $K$-fold cross-validation score as 
\begin{align*}
    \sum_{i=1}^n \sum_{j=1}^J I_{ij} W_{ij} \sum_{t=1}^T \bigg[ Y_{ijt} & - \hat{\eta}(j ,t, H_{ij}) - \{A_{ij} - \tp_{ij}(S_{ij}) \} \bigg\{ f_0(S_{ij})^T \hat\beta_0^{(-k[i])} \\
    &  + f_1(S_{ij})^T \hat\beta^{(-k[i])}_1(j) + f_2(S_{ij})^T \hat\beta^{(-k[i])}_2(t) \bigg\} \bigg] ^2,
\end{align*}
and $(K_{11}, K_{1L_1}, K_{21},..., K_{2L_2})$ are selected by minimizing this cross-validation score.

\section{Inference and Asymptotic Theory}
\label{sec:theory}

\subsection{Rate of Convergence}
We establish the asymptotic properties of the spline estimates $\hat\beta_{rl}$ when the number of subjects $n$ tends to infinity while, for each subject, the number of observations $JT$ stays fixed. We first introduce some notations. For a function $g(s)$ defined on a grid $s \in [S]$, the norm is denoted by $\| g \| = \{(1/S)\sum_{s\in [S]} g^2(s)\}^{1/2}$. For any two real sequences $\{a_n\}$ and $\{b_n\}$, we write $a_n\lesssim b_n$, or equivalently $b_n\gtrsim a_n$, if there exists an absolute constant $C$ such that $|a_n|\leq C|b_n|$ for all sufficiently large $n$. For $r = 1,2$ and $l \in [L_r]$, we say that $\hat\beta_{rl}$ is a consistent estimator of $\beta_{rl}$ if $\lim_n \|\hat\beta_{rl} - \beta_{rl}\| = 0$ holds stochastically. Let $\mathbb G_{1l}$ and $\mathbb G_{2l}$ denote the linear space spanned by $\{\phi_{lk}: k \in [K_{1l}] \}$ and $\{\psi_{lk}: k \in [K_{2l}] \}$, respectively. Let $\rho_{n}=\max_{r,l}\text{dist}(\beta_{rl}, \mathbb{G}_{rl})$, and $K_{n}=\text{max}_{r,l}K_{rl}$. The regularity conditions and technical proofs are presented in Section F of the Supplementary Material.

\begin{thm}[Consistency]
\label{thm:Consistency}
    Under Assumptions \ref{assumption:consistency}--\ref{assumption:Additive model} and regularity conditions, if $\lim K_n \log K_n/n = 0$ and $\lim_n \rho_n$ $= 0$, then $\hat\beta_{rl}$ is consistent for $r \in \{0,1,2\}$ and $l \in [L_r]$.
\end{thm}

\begin{thm}[Rate of convergence]
\label{thm:Rate of convergence}
    Under Assumptions \ref{assumption:consistency}--\ref{assumption:Additive model} and regularity conditions, if \ $\lim K_n \log K_n/n = 0$, then $\lVert\hat{\beta}_{rl}-\beta_{rl}\rVert^2=O_{p}(K_n/n + \rho_n^2)$ for $r = 0, 1, 2$ and $l \in [L_r]$.
\end{thm}

{}
$Remark \ 1. \ $ The condition $\lim K_n \log K_n/n = 0$ requires that the number of knots does not increase too quickly. This is a standard requirement in polynomial spline \citep{huang2001concave, zhang2015varying}. If specific smoothness conditions are assumed, the rate of convergence can be further simplified. For example, if the component function, $\beta_{rl}$, has bounded second derivatives, and $\mathbb G_{rl}$ is a cubic spline space,  then by Theorem 6.27 in \citet{schumaker2007spline}, $\rho_n \lesssim K_n^{-2}$. Thus, the rate of convergence in Theorem \ref{thm:Rate of convergence} becomes $O_{p}(K_n/n + K_n^{-4})$. If we let $K_n \asymp n^{1/5}$, we obtain the  optimal rate of convergence $O_p(n^{-4/5})$ as in \citet{stone1982optimal}.

\subsection{Asymptotic Variance}
\label{subsec:avar}

Let $\gamma$ denote $(\beta_0, \gamma_{1}^{(1)},...,\gamma_{L_1}^{(1)}, \gamma_{1}^{(2)},...,\gamma_{L_2}^{(2)})^T$ with $\gamma^{(1)}_{l} = (\gamma^{(1)}_{l1}, ...., \gamma^{(1)}_{lK_{1l}})$ and $\gamma^{(2)}_{l} =(\gamma^{(2)}_{l1}, ...., $ $\gamma^{(2)}_{lK_{2l}})$, $g(\gamma)$ denote the least square equation in \eqref{eq:estimating-equation}, $\dot{g}(\gamma)$ denote the derivative vector of $g(\gamma)$ with respect to $\gamma$, and $\ddot{g}(\gamma)$ denote the Hessian matrix of $g(\gamma)$ with respect to $\gamma$. We propose to estimate the uncertainty in the estimated coefficients following \citet{van1998asymptotic}.  Specifically, we estimate the variance-covariance matrix of $\hat\gamma$ by
\begin{align}
    \widehat{\var}(\hat\gamma) = \bigl\{\PP_n \ddot{g}(\hat\gamma) \bigl\}^{-1} \bigl\{\PP_n \dot{g}(\hat\gamma) \dot{g}(\hat\gamma)^T\bigl\} \bigl\{\PP_n  \ddot{g}(\hat\gamma) \bigl\} ^{-1,T}. \label{eq:var-gamma}
\end{align}
The estimated variance-covariance matrix of $\hat\beta_0$ is the upper diagonal block $L_0 \times L_0$ entry of $\widehat{\var}(\hat\gamma)$. Let $M_{[a, b]}$ denote the principal minor submatrix of $M$ obtained by deleting rows and columns outside of indices $a,a+1,\ldots,b$. Then the variance-covariance matrix of the spline estimates, $\hat\beta_1(j)$, $\hat\beta_2(t)$, are obtained as
\begin{align}
    \widehat{\var} \{\hat{\beta}_{1}(j)\} &= \Phi(j)  \widehat{\var}(\hat\gamma)_{[L_0 + 1, L_0 +\sum_{l=1}^{L_1}K_{1l}]}  \Phi(j)^T, \nonumber \\
   \widehat{\var} \{\hat{\beta}_{2}(t)\} &= \Psi(t) \widehat{\var}(\hat\gamma)_{[L_0 + \sum_{l=1}^{L_1}K_{1l} + 1, L_0 + \sum_{l=1}^{L_1}K_{1l} + \sum_{l=1}^{L_2}K_{2l}]} \Psi(t)^T,\label{eq: variance}
\end{align}
where $\Phi(j)$ is a matrix in $\mathbb R^{L_1 \times \sum_{l=1}^{L_1}K_{1l}}$ such that its $l$th row of diagonal block is B-spline basis $\{\phi_{l1}(j),...,$ $\phi_{lK_{1l}}(j)\}$, and $\Psi(t)$ is a matrix in $\mathbb R^{L_2 \times \sum_{l=1}^{L_2}K_{2l}}$ such that $l$th row of its diagonal block is B-spline basis $\{\psi_{l1}(t),...,\psi_{lK_{2l}}(t)\}$. 


Because the asymptotic variance of a Z-estimator can be anti-conservative when the sample size is small, we adopt the small sample correction technique from \citet{mancl2001} to modify the term $\PP_n \dot{g}(\hat\gamma) \dot{g}(\hat\gamma)^T$ in \eqref{eq:var-gamma}. In particular, we multiply the vector of each individual's residual by the inverse of the identity matrix minus the leverage of that individual to obtain the corrected variance. A detailed derivation is in Section G of the Supplementary Material.

An approximate $(1-\alpha)\times 100\%$ confidence interval for $\hat\beta_{1l}(j)$ or $\hat\beta_{2l}(t)$ can be constructed using \eqref{eq: variance} as
\begin{align}
    \hat\beta_{1l}(j) \pm z_{1-\alpha/2}\bigg[\widehat{\var}\big\{\hat\beta_{1l}(j)\big\}\bigg]^{1/2}, & \nonumber \\
    \hat\beta_{2l}(t) \pm z_{1-\alpha/2}\bigg[\widehat{\var}\big\{\hat\beta_{2l}(t)\big\}\bigg]^{1/2}, \label{eqn: confidence interval}
\end{align}
where $z_{1-\alpha/2}$ is the $(1-\alpha/2)$-th quantile of standard Normal distribution.

\section{Simulation Study}
\label{sec:simulation}

We generated data for a hypothetical MRT with $J = 30$ and $T = 20$. For participant $i$ and decision point $j$, an exogeneous binary covariate $S_{ij}$ is generated from $\text{Bernoulli}(0.2)$, and an exogeneous continuous covariate $Z_{ij}$ is generated from $\text{Normal}(0,1)$. The binary treatment $A_{ij}, j\in[J]$ is generated from $\text{Bernoulli}(0.4)$. In the time-varying functions we scaled both time indices to take values in $[0,1]$: let $\tilde{j} = j/J$ and $\tilde{t} = t/T$, and we generated the proximal outcome as
\begin{align*}
    Y_{ijt} &= A_{ij}\big\{\beta_0 + \beta_1(\tilde{j}) + \beta_{21}(\tilde{t}) +  S_{ij}\beta_{22}(\tilde{t}) \big\} \\ 
    &\ \ \ + \alpha_0 + \alpha_{11}(\tilde{j}) + S_{ij}\alpha_{12}(\tilde{j}) + \alpha_{21}(\tilde{t}) + Z_{ij}\alpha_{22}(\tilde{t}) + e_{ijt},
\end{align*}
where the $e_{ijt}$'s are marginally $\text{Normal}(0,1)$, independent across different $i$ or $j$, and for the $i$-th individual at the $j$-th decision point $\text{Corr}(e_{ijt}, e_{ijs}) = 0.4$ for $t \neq s$. The causal excursion effect is thus
\begin{align}
    \cee(j, t, S_{ij}) = \beta_0 + \beta_1(\tilde{j}) + \beta_{21}(\tilde{t}) +  S_{ij}\beta_{22}(\tilde{t}). \label{eq:dgm-cee}
\end{align}

\begin{figure}
    \centering
    \includegraphics[scale = 0.38]{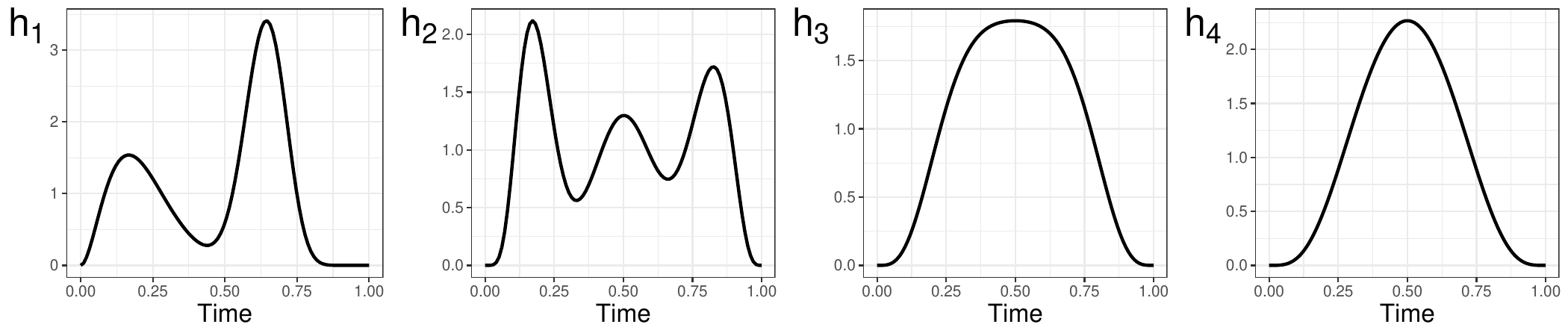}
    \caption{Functions $h_1$, $h_2$, $h_3$, and $h_4$ used in the simulations.}
   \label{fig:shape of the functions}
\end{figure}

We set $\alpha_0 = 0.35$ and $\beta_0 = 0.7$. For the time-varying functions, we constructed two relatively complex functions $h_1(\cdot), h_2(\cdot)$ and two relatively simple functions $h_3(\cdot), h_4(\cdot)$, all defined on $[0,1]$ and adapted from \citet{wahba1983bayesian}: 
\begin{align*}
    h_1(\cdot) &= \frac{6}{10}g_{30,17}(\cdot) + \frac{4}{10}g_{3,11}(\cdot), 
    &&h_2(\cdot) = \frac{1}{3}g_{20,5}(\cdot) + \frac{1}{3}g_{12,12}(\cdot) + \frac{1}{3}g_{7, 30}(\cdot) \\
    h_3(\cdot) &= \frac{1}{3}g_{10,5}(\cdot) + \frac{1}{3}g_{7,7}(\cdot) + \frac{1}{3}g_{5,10}(\cdot), 
    &&h_4(\cdot) = \frac{1}{3}g_{8,5}(\cdot) + \frac{1}{3}g_{7,7}(\cdot) + \frac{1}{3}g_{5,8}(\cdot). 
\end{align*}
where $g_{a,b}(\cdot)$ denotes the probability density function of a Beta distribution with shape parameters $a,b$ (Figure \ref{fig:shape of the functions}). We considered the following simulation settings :
\begin{itemize}
    \item (Truth being complex functions.) $\beta_1 = h_1$, $\beta_{21} = h_1$, $\beta_{22} = h_2$, $\alpha_{11} = h_1$, $\alpha_{12} = h_3$, $\alpha_{21} = h_1$, $\alpha_{22} = h_2$. 
    \item (Truth being simple functions.) $\beta_1 = h_3$, $\beta_{21} = h_3$, $\beta_{22} = h_4$, $\alpha_{11} = h_3$, $\alpha_{12} = h_3$, $\alpha_{21} = h_3$, $\alpha_{22} = h_2$.
\end{itemize}

In the first stage of the estimator, we used the working model for expected outcome under no treatment: $\eta(j,t,H_{ij}) = \alpha_0 + \alpha_1(\tilde{j}) + \alpha_{21}(\tilde{t}) + S_{ij}\alpha_{22}(\tilde{t})$. We misspecified the working model by omitting $Z_{ij} \alpha_{22}(\tilde{t})$ to assess the robustness of the estimator. The causal excursion effect is correctly modeled as \eqref{eq:dgm-cee}. We used cubic splines with evenly spaced knots between $[0,1]$ as the nonparametric function approximators for $\beta_1(\cdot)$, $\beta_{21}(\cdot)$, $\beta_{22}(\cdot)$, $\alpha_1(\cdot)$, $\alpha_{21}(\cdot)$, and $\alpha_{22}(\cdot)$. The numbers of knots were selected for each simulated data set using 10-fold cross-validation and optimized using a one-pass coordinate search for computational efficiency.

\begin{table}
    \centering
    \caption{MSE and coverage probability of the estimated intercept in CEE ($\hat\beta_0$) under complex (top half) and simple (bottom half) settings in the simulation.}
    \label{tab: MSE and CP of beta0}
    \begin{tabular}{c c c c} 
         \toprule
        Simulation setting & Sample size & MSE & Coverage Probability\\ 
        \midrule
        \multirow{3}{*}{Complex functions} & n = 20 & 0.061 & 0.911  \\ 
        & n = 50 & 0.017 & 0.915 \\
        & n = 100 & 0.007 & 0.938 \\
        \midrule
        \multirow{3}{*}{Simple  functions} & n = 20 & 0.057 & 0.911 \\
        & n = 50 & 0.016 & 0.913\\
        & n = 100 & 0.007& 0.938\\
         \bottomrule
    \end{tabular}
\end{table}

\begin{figure}
    \centering
    \includegraphics[scale = 0.37]{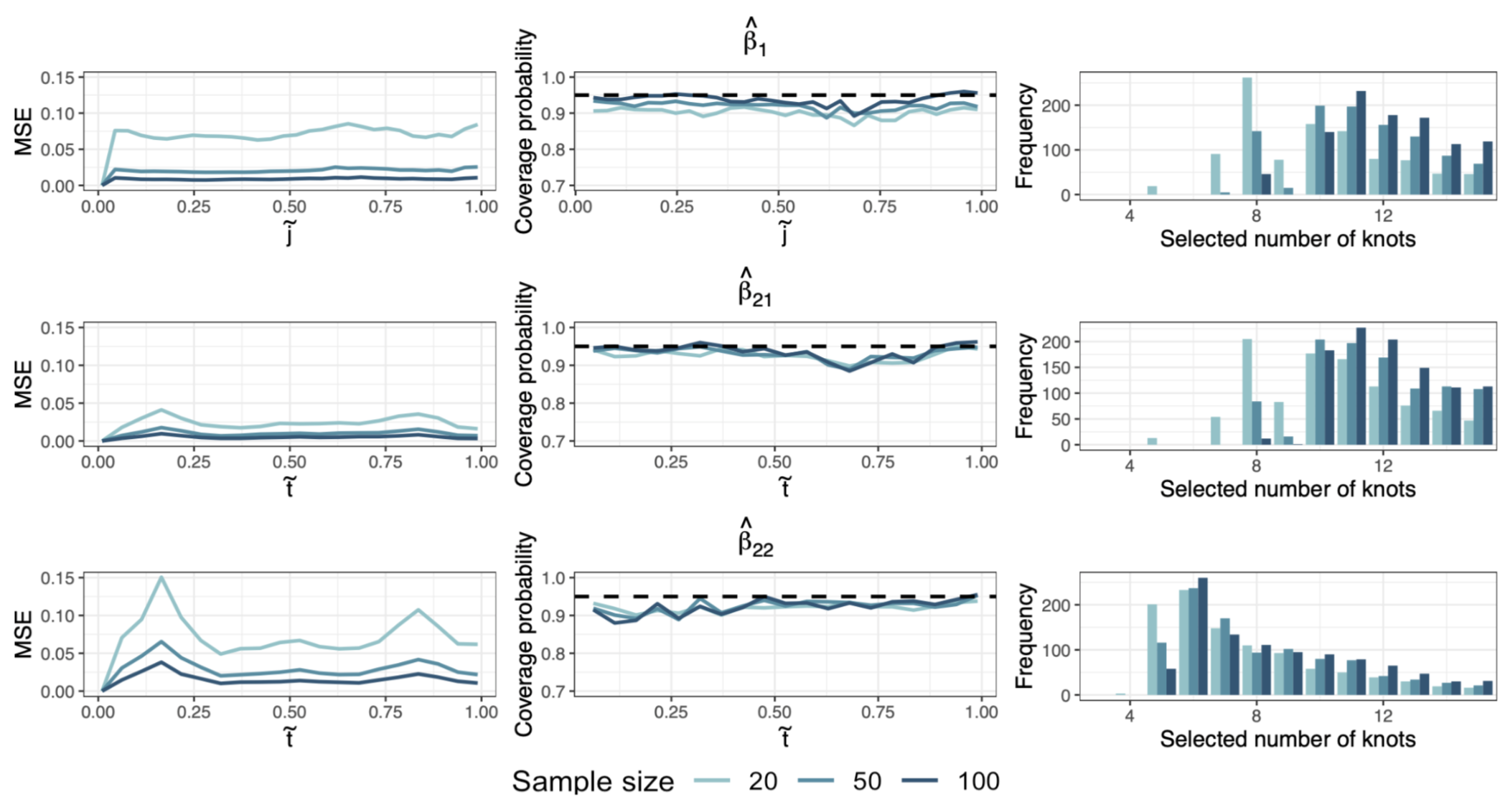}
    \caption{Mean squared error (left), coverage probability of 95\% confidence interval (center), and selected number of knots (right) for $\hat\beta_1(\tilde{j})$, $\hat\beta_{21}(\tilde{t})$, and $\hat\beta_{22}(\tilde{t})$ under the complex function setting in the simulation.}
   \label{fig:complex simulation results for varying N}
\end{figure}

We considered sample sizes $n = 20, 50, 100$ and conducted each simulation for 1000 replicates. For the setting with complex time-varying functions, as the sample size increases, the mean squared error (MSE) of the estimated intercept in CEE ($\hat\beta_0$) decreases (Table \ref{tab: MSE and CP of beta0}, top half), and the point-wise MSE of $\hat\beta_1(\tilde{j})$, $\hat\beta_{21}(\tilde{t})$, and $\hat\beta_{22}(\tilde{t})$ decreases (Figure \ref{fig:complex simulation results for varying N}, left column). This verifies the consistency result (Theorem \ref{thm:Consistency}). In addition, the coverage probability of 95\% confidence interval is close to the nominal level even with a small sample size such as $n=20$ (Figure \ref{fig:complex simulation results for varying N}, middle column). Lastly, as the sample size increases, the number of selected knots increases slightly (Figure \ref{fig:complex simulation results for varying N}, right column). 

\begin{figure}
    \centering
    \includegraphics[scale = 0.34]{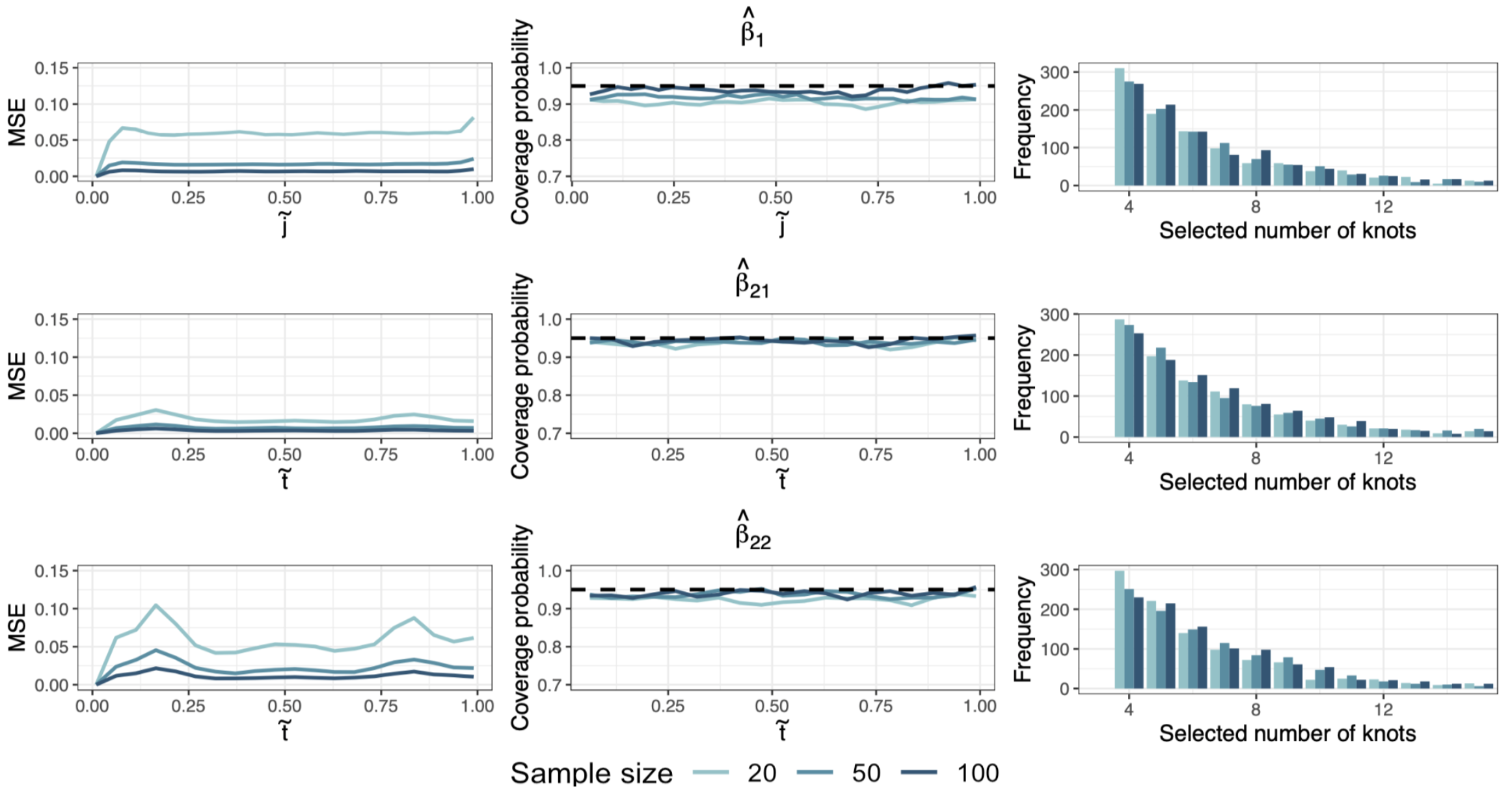}
    \caption{Mean squared error (left), coverage probability of 95\% confidence interval (center), and selected number of knots (right) for $\hat\beta_1(\tilde{j})$, $\hat\beta_{21}(\tilde{t})$, and $\hat\beta_{22}(\tilde{t})$ under the simple function setting in the simulation.}
   \label{fig:simple simulation results for varying N}
\end{figure}

For the setting with simple time-varying functions, similar observations can be made regarding the decreasing MSE and close-to-nominal coverage probability (Table \ref{tab: MSE and CP of beta0}, bottom half; Figure \ref{fig:simple simulation results for varying N}).  In addition, comparing the right column between Figures \ref{fig:complex simulation results for varying N} and \ref{fig:simple simulation results for varying N} shows that the selected number of knots is larger when the time-varying functions are complex.

\section{Application}
\label{sec:application}

We assessed the effect of activity suggestions in the HeartSteps MRT on participants' minute-level step count using the proposed method. 37 participants were included in the analysis with 7629 decision points. Even though seven participants had less than 210 decision points of data (with reasons detailed in \citet{klasnja2018efficacy}), the same number of knots and knot locations were used for everyone. 

Because an activity suggestion may either be a walking suggestion or an anti-sedentary suggestion, we used a generalized version of the proposed model (detailed in Section D of the Supplementary Material to accommodate categorical treatments. Let $A_{ij}$ take three values: $0$ if no push notification was delivered (with probability 0.4), $1$ if an activity suggestion was delivered (with probability 0.3), and $2$ if an anti-sedentary suggestion was delivered (with probability 0.3). Let the proximal outcome $Y_{ijt}$ be the log-transformed step count during the $t$-th minute following decision point $j$ of participant $i$. Let $I_{ij}$ denote participant $i$'s availability at decision point $j$: participants were considered unavailable at a decision point, and thus no treatment would be delivered, if they were currently driving, had manually turned off the intervention, were exercising 90 seconds before the decision point, or did not have an active internet connection \citep{seewald2019practical}. Among the 7629 decision points, 6137 were available.

The original primary analysis of HeartSteps focused on a 30-minute aggregate step count proximal outcome \citep{klasnja2018efficacy}. To assess how soon an individual responds to the suggestion, we consider the 60-minute time window following each decision point, i.e., $T = 60$. This also helps to assess whether the 30-minute window was sufficiently wide to capture the effect of the activity suggestions in the original analysis.

We first assessed how the CEE of the walking suggestions and the anti-sedentary suggestions varied within the 60-minute window and throughout the study. Then we investigated how the CEE is moderated by three time-varying variables: location, weekday/weekend, and recent activity level. Those were among the variables used to tailor the content of the suggestions \citep{smith2017design}.

Lastly, we mimic what a typical MRT analysis would be by examining aggregate outcomes, which further highlights the added value of examining the more granular longitudinal functional outcome.

\subsection{Marginal Causal Excursion Effect}
\label{subsec: marginal analysis}

We considered the following CEE model for the walking suggestion ($A_{ij} = 1$) and the anti-sedentary suggestion ($A_{ij} = 2$), compared to no suggestion ($A_{ij} = 0$):
\begin{align}
    \EE\{Y_{ijt}(\bA_{i,j-1}, m) - Y_{ijt}(\bA_{i,j-1}, 0) \mid I_{ij} = 1 \} =  \beta_0^{(m)} + \beta_1^{(m)}(j) + \beta_2^{(m)}(t), \quad m = 1, 2. \label{eq:marginal-analysis-cee-walking}
\end{align}  
Here, $\beta_0^{(m)}$ represents the effect of the corresponding suggestion ($m=1$ for walking, $m=2$ for anti-sedentary), compared to no suggestion, on the log-transformed step count at the first minute following the first decision point (recall the identifiability constraints $\beta_1^{(m)}(1) = \beta_2^{(m)}(1) = 0$). $\beta_1^{(m)}(j)$ captures how the effect of the corresponding suggestion varies over decision points. $\beta_2^{(m)}(t)$ captures how the effect of the corresponding suggestion varies over 60 minutes following a decision point.

\begin{figure}
    \centering
    \includegraphics[scale = 0.42]{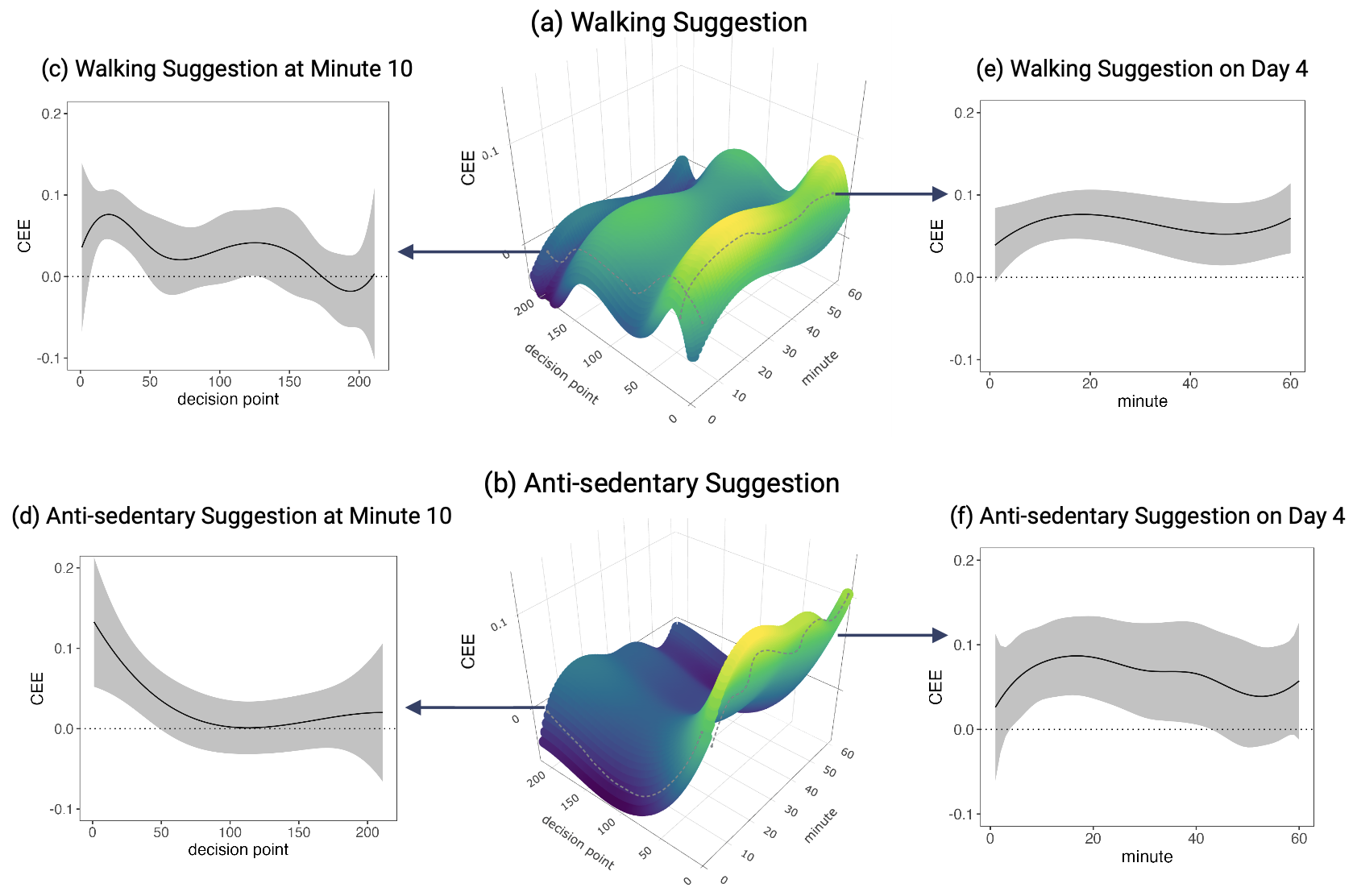}
    \caption{Surface plots for CEE of (a) walking suggestions and (b) anti-sedentary suggestions over decision points and minutes from the analysis in Section \ref{subsec: marginal analysis}. A lighter color represents a larger effect. Panels (c) and (d) are slices of the surface plots at minute 10 after receiving the notification, illustrating the CEE over the course of the study period; panels (e) and (f) are slices of the surface plots at decision point 20 (day 4), illustrating the CEE over 60 minutes following the decision point.}
    \label{fig:DA-surface plot}
\end{figure}

The Stage 1 nuisance function $\eta(j, t, H_{ij})$ was fitted using a generalized additive model with penalized splines, implemented by the \texttt{gam} function in the \texttt{mgcv} R package \citep{wood2017generalized}. We included as features the minute index $t$, the decision point index $j$, location (an indicator for being at home/work or not), the weekday indicator, recent activity level (an indicator of whether an individual took less than 56 steps in the 30-minute window prior to a decision point), and individual's log-transformed cumulative step counts in a 30-minute window following the previous decision point. The Stage 2 functional parameters $\beta_1^{(m)}(j)$ and $\beta_2^{(m)}(t)$ were modeled using cubic spline. The number of knots was selected through a one-pass coordinate search using 10-fold cross-validation, and the selected number of interior knots for $\hat\beta_{1}^{(1)}(j), \hat\beta_{2}^{(1)}(t)$, $\hat\beta_{1}^{(2)}(j)$, and $\hat\beta_{2}^{(2)}(t)$ were 3, 0, 0, and 5, respectively. 

Figure \ref{fig:DA-surface plot} shows the trajectories of CEE over decision points and minutes for the walking suggestion and the anti-sedentary suggestion. For both types of suggestions, peaks of the effects (the yellow region in Figure \ref{fig:DA-surface plot}(a) and (b)) occurred at the beginning of the study and within 30 minutes following a decision point. Figure \ref{fig:DA-surface plot}(c) and (d) both show a decreasing CEE over decision points where the largest effect of both types of suggestions occurred early on in the study. In addition, we observed that compared with walking suggestions, anti-sedentary suggestions started with a greater effect but the effect declined much faster over decision points. When examining the effect over minutes at a fixed decision point (the plots showing day 4), significantly positive effects occurred around minute 20 for both types of suggestions, with the walking suggestion having a positive effect over a wider window up to minute 40 (Figure \ref{fig:DA-surface plot}(e) and (f)). This observation aligns with the intervention design in that the suggested activities were brief, with the walking suggestions expected to take longer to complete than the anti-sedentary suggestions. The delay between receiving (minute 0) and acting upon the suggestion (the peak) is attributed to the \textit{in situ} nature of the suggestions as they were integrated into the participants' daily lives.

\subsection{Moderated Causal Excursion Effect}
\label{subsec: moderation analysis}

We considered the following CEE model for the effect moderated by a single time-varying covariate, $S_{ij}$:
\begin{align}
    & ~~~~ \EE\{Y_{ijt}(\bA_{i,j-1}, k) - Y_{ijt}(\bA_{i,j-1}, 0) \mid I_{ij} = 1, S_{ij} \} \nonumber \\ & =  \beta_{01}^{(k)} + \beta_{02}^{(k)}S_{ij} + \beta_{11}^{(k)}(j) + S_{ij} \beta_{12}^{(k)}(j) + \beta_{21}^{(k)}(t) + S_{ij} \beta_{22}^{(k)}(t), \quad k = 1, 2. \label{eq:mod analysis-cee-walking}
\end{align} 
We conducted three moderation analyses: (i) $S_{ij} = \loc_{ij}$, which equals 1 if the participant was at home or work and 0 otherwise; (ii) $S_{ij} = \weekday_{ij}$, which equals 1 if the $j$-th decision point was on a weekday and 0 otherwise; and (iii) $S_{ij} = \sedentary_{ij}$, which equals 1 if the participant was sedentary (less than 56 steps in the 30 minutes before $j$-th decision point), using the median step count as the threshold.

The nuisance function $\eta(j, t, H_{ij})$ in Stage 1 and the functional $\beta$'s in Stage 2 were fitted using the same approach as the marginal CEE analysis in Section \ref{subsec: marginal analysis}, except that the corresponding moderator was included as an additional feature in Stage 1. The selected number of interior knots for the $\beta$'s are in Table \ref{tab: knots number selected for moderator analysis}. To better interpret the results and answer the research questions raised at the end of Section \ref{sec: HeartSteps MRT Data}, we calculated and presented the treatment effects conditional on various values of moderators.

\begin{table}
\centering
    \caption{Selected number of knots for each functions in the moderator analysis.}
    \label{tab: knots number selected for moderator analysis}
    \begin{tabular}{c c c c c c c c c} 
         \toprule
         Moderator & $\beta_{11}^{(1)}(j)$ & $\beta^{(1)}_{12}(j)$ & $\beta^{(1)}_{21}(t)$ & $\beta^{(1)}_{22}(t)$ & $\beta_{11}^{(2)}(j)$ & $\beta^{(2)}_{12}(j)$ & $\beta^{(2)}_{21}(t)$ & $\beta^{(2)}_{22}(t)$\\ 
        \midrule
        Weekday/Weekend & 2 & 3 & 0 & 5 & 0 & 4 & 1 & 7 \\
        Location & 3 & 1 & 0 & 7 & 0 & 7 & 3 & 4 \\
        Sedentary & 1& 4 &2 &4 & 1 &6 &3 &6\\
         \bottomrule
    \end{tabular}
\end{table}

\begin{figure}
    \centering
    \includegraphics[scale = 0.51]{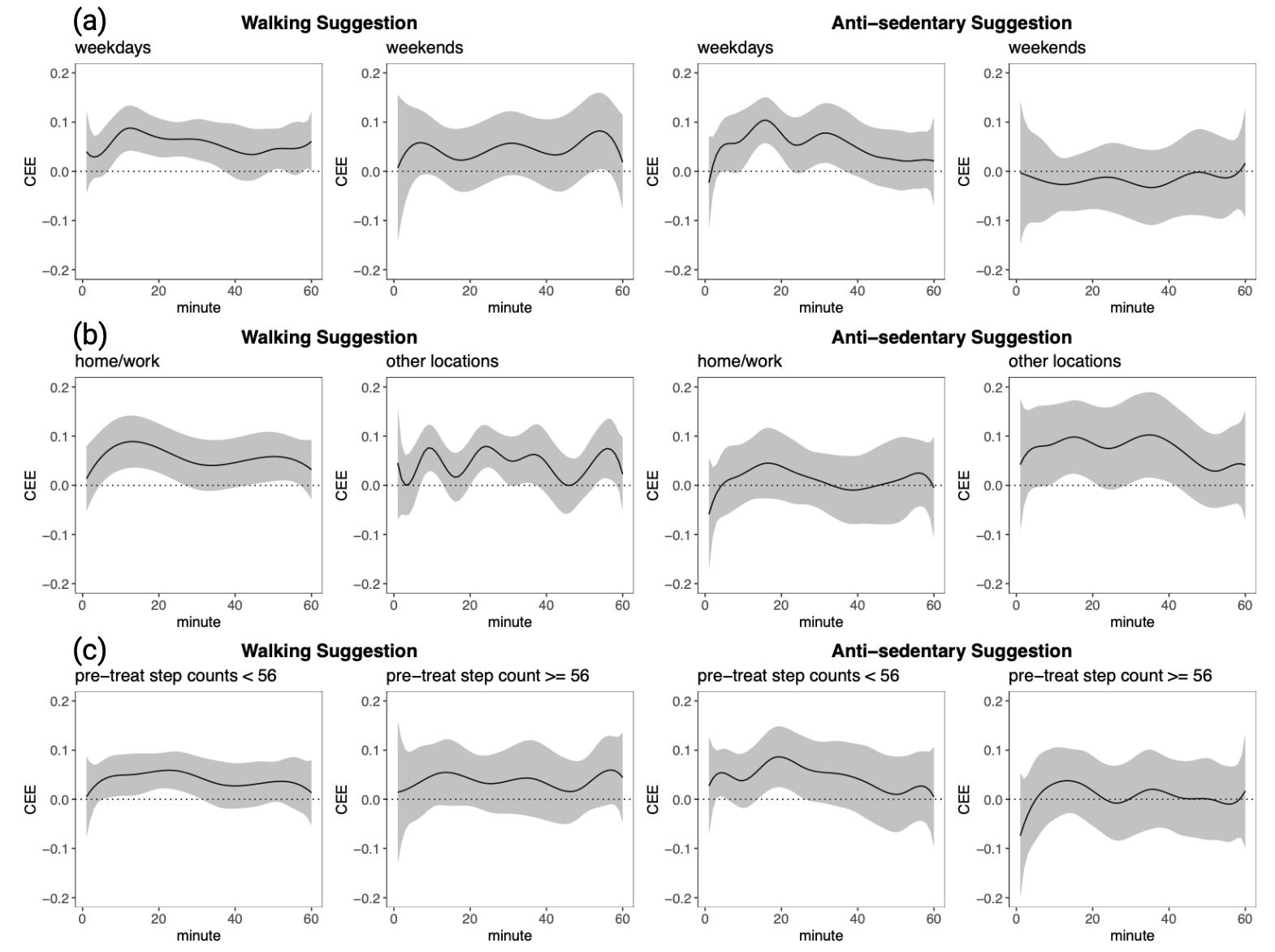}
    \caption{Estimated CEE of walking (left columns) and anti-sedentary (right columns) suggestions (compared to no suggestion) on minute-level step counts over a 60-minute period following the 35-th decision point (day 7), moderated by  (a) location, (b) weekday/weekend, and (c) 30-min pre-treatment total step count.}
    \label{fig: three moderators by min}
\end{figure}

\begin{figure}
    \centering
    \includegraphics[scale = 0.51]{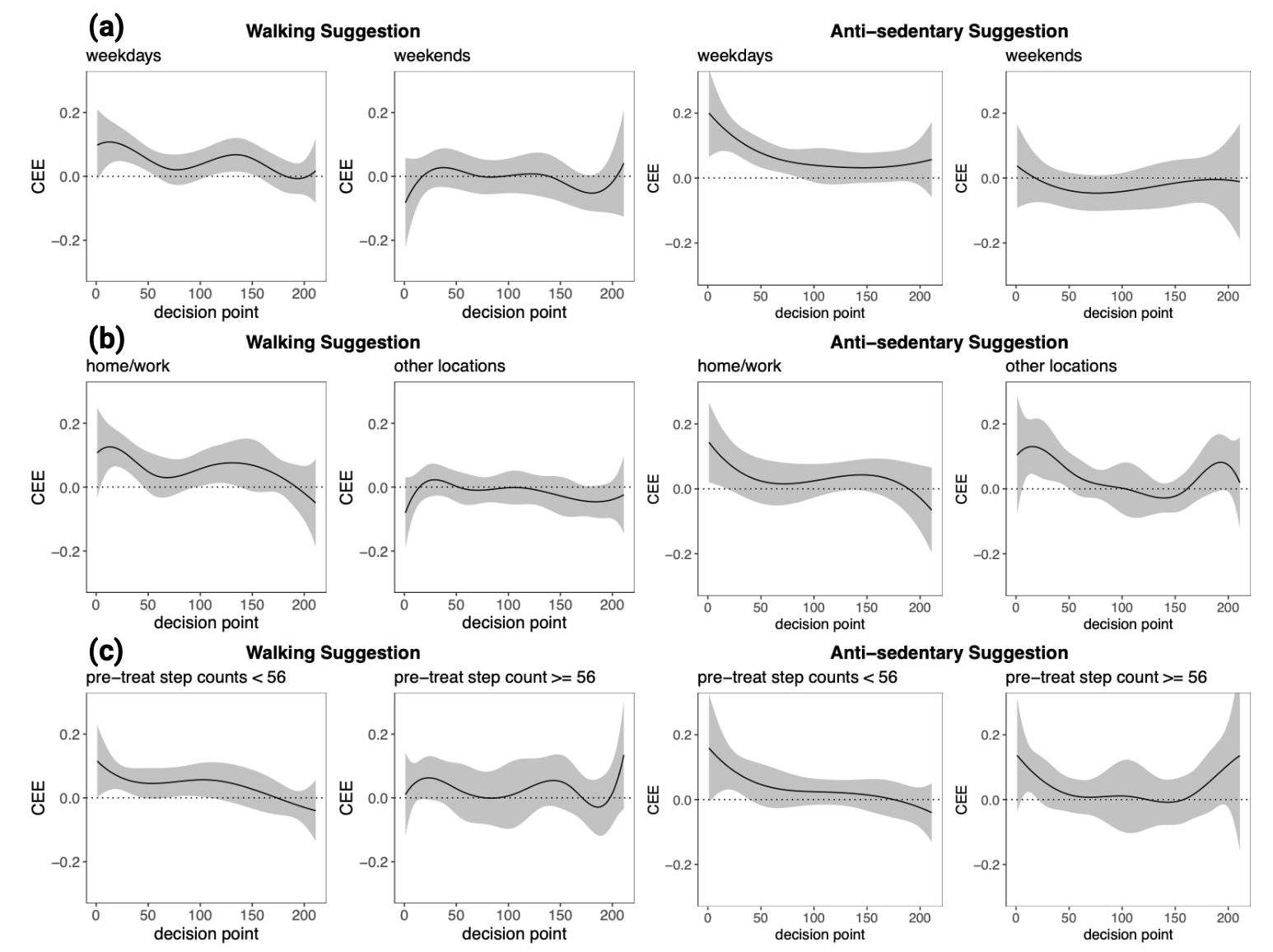}
    \caption{Estimated CEE of walking (left columns) and anti-sedentary (right rolumns) suggestions (compared to no suggestion) on the step count in the 15-th minute following each of the 210 decision point over the course of the study, moderated by  (a) location, (b) weekday/weekend, and (c) 30-min pre-treatment total step count.}
    \label{fig:three moderators by dec}
\end{figure}

Figure \ref{fig: three moderators by min} shows the CEE profile over the 60 minutes following decision point $j=35$ (day 7) from these moderation analyses. Figures for other decision points are included in Section A of the Supplementary Material, and they are vertical shifts of Figure \ref{fig: three moderators by min} due to the additive nature of the model. Figure \ref{fig: three moderators by min}(a) shows that on weekdays, both the walking suggestion and the anti-sedentary suggestion had a significant positive effect around minutes 10-20, whereas on weekends neither suggestion had a positive effect. This indicates that structured environments may facilitate more routine responses; this could also be due to the suggestions being better tailored on weekdays. Figure \ref{fig: three moderators by min}(b) shows that only the walking suggestion exhibits a significantly positive effect in a narrow window around the peak of 10-20 minutes. Figure \ref{fig: three moderators by min}(c) shows that both suggestions were more effective when the participant was sedentary (compared to not) during the prior 20-40 minutes. The walking suggestion shows a significantly positive effect lasting for 40 minutes with a peak around minute 10-20 when $\sedentary_{ij} = 1$, and this positive effect persisted through the first quarter of the study (see Figure \ref{fig:three moderators by dec}(c)).
This implies that it was easier to nudge the participants when they were in a sedentary state.

Figure \ref{fig:three moderators by dec} shows the CEE profile over decision points at minute 15 from these moderation analyses. Additional figures for other minutes are in the Supplementary Material, and they are vertical shifts of Figure \ref{fig:three moderators by dec}. When the day was a weekday, when the participant was at home/work, or when the participant was recently sedentary, suggestions were effective in promoting physical activity, especially during the first quarter of the study (up to around decision point 50). Comparing the anti-sedentary suggestion with the walking suggestion, the former started with a greater effect but had a faster decline over decision points, and thus its effect became insignificant earlier in the study. This indicates that participants may have become habituated to anti-sedentary suggestions particularly quickly, which may be alleviated in future studies by more variability in the suggestion content.

\subsection{Causal Excursion Effect on Aggregate Outcome}
\label{subsec: additional analysis: aggregate outcome}

\begin{table}
\centering
    \scriptsize
  \begin{sideways}
  \begin{minipage}{22cm} 
\caption{Coefficient Estimate, SE, and p-value for the two estimands stated in Section \ref{subsec: additional analysis: aggregate outcome}: (a) fully marginal effect. (b) moderated effect. ``Walking'' indicates the coefficient of the treatment variable for walking suggestion; ``walking $\times$ decision point'' indicates the coefficient of the treatment variable for walking suggestion interact with time; ``anti-sedentary'' indicates the coefficient of the treatment variable for anti-sedentary suggestion; ``anti-sedentary $\times$ decision point'' indicates the coefficient of the treatment variable for anti-sedentary suggestion interact with time.}
        \label{tab: WCLS no moderator}
       \centering
  \begin{tabular}{lcc ccc ccc ccc ccc} 
            \toprule
            &  &
            \multicolumn{3}{c}{Walking} &
            \multicolumn{3}{c}{Walking $\times$ decision point} &
            \multicolumn{3}{c}{Anti-sedentary} &
            \multicolumn{3}{c}{Anti-sedentary $\times$ decision point} \\
            \cmidrule(lr){3-5} \cmidrule(lr){6-8} \cmidrule(lr){9-11} \cmidrule(lr){12-14}
            Outcome & Estimand & Est. & SE & $\text{p-value}$ & Est. & SE & $\text{p-value}$ & Est. & SE & $\text{p-value}$ & Est. & SE & $\text{p-value}$ \\
            \midrule
            \multirow{2}{*}{30-min} & (a) & 0.200 & 0.087 & 0.029 
            &  &  & 
            & 0.017 & 0.090 & 0.848
            &  &  &  \\
            & (b) & 0.730 & 0.150 & <0.001
            & -0.005 & 0.001 & <0.001
            & 0.238 & 0.198 & 0.241
            & -0.002 & 0.002 & 0.232 \\
            \multirow{2}{*}{40-min} & (a) & 0.225 & 0.094 & 0.024
            &  &  & 
            & 0.047 & 0.103 & 0.651 
            &  &  & \\
            & (b) & 0.782 & 0.157 & <0.001
            & -0.005 & 0.001 & <0.001
            & 0.269 & 0.214 & 0.209 
            & -0.002 & 0.002 & 0.247 \\
            \multirow{2}{*}{60-min} & (a) & 0.135 & 0.097 & 0.174
            &  &  & 
            & -0.001 & 0.099 & 0.995
            &  &  &  \\
            & (b) & 0.565 & 0.166 & 0.002
            & -0.004 & 0.001 & 0.010
            & 0.294 & 0.186 & 0.126
            & -0.003 & 0.002 & 0.089 \\
            \bottomrule
        \end{tabular}
  \end{minipage} 
  \end{sideways}
 \end{table}

\citet{klasnja2018efficacy} found that walking suggestions had a positive effect on an aggregate proximal outcome, specifically the total step count within 30 minutes following a decision point. However, our double indices analyses in Section \ref{subsec: marginal analysis} suggest that this positive effect extends beyond the 30-minute window. This motivated us to replicate and extend the work of \citet{klasnja2018efficacy}, which focused on the 30-minute step count outcome. In this subsection, we explored aggregate outcomes over a longer time window to determine whether the 40-minute step count, as indicated by our analyses, shows a greater effect compared to the 30-minute total step count.

Specifically, we considered three aggregate outcomes: the total step count in a 30-minute, 40-minute, or 60-minute window following a decision point. For each of the outcomes, we applied the same analysis method used in \citet{klasnja2018efficacy} and considered two estimands: (a) a fully marginal effect of the suggestion averaged over all decision points and all participants; and (b) a moderated effect with the decision point index being the effect modifier. We included the same control variables as the previous analyses for fitting a nuisance parameter in the weighted and centered least squares ($g_t(H_t)^T \alpha$ in \citep{boruvka2018assessing}).

The analysis results for the three aggregate outcomes are displayed in Table \ref{tab: WCLS no moderator}. The same qualitative conclusions as in \citet{klasnja2018efficacy} hold for the 40-minute and the 60-minute total step count: Averaging over study days, delivering a walking suggestion versus providing no suggestion increased the average aggregate step count significantly, whereas anti-sedentary suggestion has no significant effect. In addition, the effects deteriorate over the course of the study. Quantitatively, the magnitude of the causal effect of the suggestions on the 40-minute total step counts is greater than that on the 30-minute or the 60-minute total step counts. This suggests that a time window of 40 minutes, which was discovered through our proposed method (Section \ref{subsec: marginal analysis} and \ref{subsec: moderation analysis}), may be more appropriate for capturing the impact of the activity suggestions on step count.

Additional moderation analysis results for the three aggregate outcomes can be found in Section A of the Supplementary Material. They also suggest the superiority of a 40-minute time window in capturing the impact of activity suggestions on step count.

\subsection{Sensitivity Analysis}

The Jawbone wristband trackers used in HeartSteps were designed to only record positive step counts \citep{seewald2019practical}. Therefore, it was difficult to determine if a minute with a recorded zero step count was a minute with truly zero steps or a minute when the tracker was not worn. The analyses in Sections \ref{subsec: marginal analysis} and \ref{subsec: moderation analysis} assumed that zero-step minutes as no movement. To assess the sensitivity of our results to these assumptions, we imputed the zero step counts using the step count recorded via Google Fit. Google Fit step count was smartphone-based and thus less reliable than the wristband tracker, but it may nonetheless supplement the step count information when the wristband tracker was not worn (see Figure 4 of the Supplementary Material).

With the zero step counts imputed with Google Fit data and the nonzero step counts unchanged, we conducted the same analyses as in Sections \ref{subsec: marginal analysis} and \ref{subsec: moderation analysis}. The conclusions in these sections still hold with the imputed data. Figures showing the sensitivity analysis results using the imputed data can be found in Section C of the Supplementary Material.

\section{Discussion}
\label{sec:discussion}

We proposed a time-varying effect model on the causal excursion effect for longitudinal functional outcomes. To our knowledge, this is the first study of applying nonparametric causal models on longitudinal functional data. Our model incorporated double time indices in an additive model for the causal effects, and the two-stage estimator we proposed demonstrates robustness against misspecification of a high-dimensional nuisance parameter, which is validated by both theoretical analyses and simulation studies.

Our analysis of the HeartSteps MRT data revealed new insights regarding how participants responded to the activity suggestions and how such response was moderated by time-varying contexts. Here we summarize the answers to the research questions posed in Section \ref{sec: HeartSteps MRT Data}. Our analysis found that the participants started to respond around 10 minutes after receiving the suggestion and the effect tapered off around the 30-minute mark, with the walking suggestion having a longer time window of positive effect compared to the anti-sedentary suggestion. The pattern of the time-varying effect of the suggestions depended greatly on the moderators we considered, especially whether a participant was recently sedentary and whether the day was a weekday. Generally speaking, an individual in a more structured environment who was recently sedentary would be more responsive to the activity suggestions, especially the walking suggestions. In terms of the time-varying effect pattern throughout the study, the effect of the anti-sedentary suggestion, compared to the walking suggestion, was greater at the beginning of the study but declined faster and became ineffective sooner.

Our methodology and analysis come with limitations. First, because the wristband tracker used in HeartSteps did not distinguish a minute with truly zero step counts from a minute when the tracker was not worn, both minutes would yield an apparent zero step count in the data set. This limitation was alleviated by our sensitivity analysis using supplementary data collected via the smartphone. A future direction is to systematically study the handling of missing data using multiple data sources, especially this type of sensor-collected data in which even the missingness indicator is not recorded.

Second, the minute-level step counts were zero-inflated (Figure \ref{fig:exploratory plot_heatmap}). Even though our causal model does not make assumptions about the outcome distribution and thus yields valid causal estimates for zero-inflated outcomes, future research may take into account the zero-inflated nature in either or both of the causal estimand definition and the estimation of nuisance parameters to improve the interpretability and efficiency. Promising directions include two-part models for zero-inflated outcome \citep{lambert1992zero,mullahy1986specification}, principal stratification-based causal effects \citep{frangakis2002principal,staub2014causal,lee2017extensive}, nonzero membership-based mediation analysis \citep{wang2012estimation}, and using zero-inflated models for nuisance parameter estimation \citep{yu2023multiplicative}. Given the controversies in the causal effect definitions for zero-inflated outcomes \citep{angrist2001estimation}, a comprehensive study of MRT analyses with zero-inflated outcomes is left for future work.

\section*{Supplementary Material}
We show additional figures and tables for the application that can supplement our analysis but were not included in the main paper, and present the details of sensitivity analysis results. We demonstrate the generalization of our method to categorical treatments. We also show the proofs of Theorem 1 and 2. The code for implementing the proposed method in simulations and data application can be found on \url{https://github.com/jiaxin4/MobileHealth}

\section*{Acknowledgements}
The authors would like to thank Susan Murphy, Ashley Walton, and Justin Zhu for their helpful feedback. The authors appreciate the financial support of UC Irvine School of Information and Computer Sciences through the ICS Research Award.

\bibliographystyle{apalike}
\bibliography{ref}


\end{document}